\documentclass[aps,prc,onecolumn,superscriptaddress]{revtex4}
\usepackage[latin1]{inputenc}
\usepackage{bm}
\usepackage{epsfig}
\usepackage{amsthm}
\usepackage{amsfonts}
\usepackage{float}
\usepackage{amsmath,amssymb}
\usepackage{color}
\usepackage{slashed}
\usepackage{relsize}
\usepackage[toc,page]{appendix} 
\usepackage{ulem}
\usepackage{cancel}
\usepackage{tikz}
\usepackage{hyperref}
\usepackage{fontenc}
\usepackage{graphicx}
\usepackage{dcolumn}
\usepackage{bm}
\newcommand{\be}{\begin{equation}} 
\newcommand{\ee}{\end{equation}} 
\newcommand{\bea}{\begin{eqnarray}} 
\newcommand{\eea}{\end{eqnarray}} 
\newcommand{\nn}{\nonumber} 
\newcommand{\mintedim}[2]{{\int\kern-0.50em\mbox{{\small$\mathop{\frac{\mbox{{\small${\rm d^{#2}}\vect{#1}$}}}{\mbox{{\small$(2\pi)^{#2}$}}}}$}}\ }} 
\newcommand{\inteonedim}[1]{{\int_0^\infty\kern-1em\mbox{{\small${\rm d}{#1}$}}}} 
\newcommand{\vect}[1]{\bm{#1}} 

\begin{document}
\title{
\textcolor{black} {A Tsallis-like effective exponential delay discounting model and its implications}
}
\author{Trambak Bhattacharyya}
\email{bhattacharyya@theor.jinr.ru (corresponding author)}
\affiliation{Bogoliubov Laboratory of Theoretical Physics, Joint Institute for Nuclear Research, Dubna, 141980, \\
Moscow Region, Russia}

\author{Shanu Shukla}
\email{shanu.shukla11@gmail.com}
\affiliation{Indian Institute of Management Indore, Rau-Pithampur Road, Indore, Madhya Pradesh, India, 453556 }

\author{Ranu Pandey}
\email{ranupandey@jssaten.ac.in}
\affiliation{JSS Academy of Technical Education, Noida, India, 201301}

\begin{abstract} 
We propose an effective exponential model of delay discounting considering fluctuation in impulsivity. This model is seen
to be dual to the two-parameter Tsallis model of delay discounting proposed by Takahashi in 2007 (Ref.~\cite{takahashi1}). We
demonstrate that the parameters of the model can be related to the mean and the variance of impulsivity. Our findings 
provide an intuitive way to explore the origin of the Tsallis distribution in a social system. 
\\
\\
Keywords: Tsallis statistics, delay discounting model, impulsivity fluctuation
\end{abstract}
\maketitle
%

\section{Introduction}

Delay discounting is associated with the evolution of the value of a reward with delay to its receipt. For example, money
of an amount $x$ received immediately is more `valuable' than the amount received after a delay time $t$. However, when
some features about a situation remain unspecified, uncertainties in decision arise and making a choice becomes more difficult.
One such example is choosing between USD 50 right now and USD 70 in a year. In this case, one has to choose between
smaller-sooner rewards and larger-later rewards. One example of choosing a smaller-sooner reward is a student unprepared for an exam going to a party the night 
before the test. The immediate reward brought out of this action appears to be more desirable than a delayed reward (which may be of greater
value in terms of career, earning and so on). Choosing the smaller-sooner reward is termed as an impulsive action by the behavioral analysts
 \cite{ikuk}. This implies recognizing more distant rewards as of lesser value. For a variety of species, reward types and sample, the value of a reward
is discounted over time \cite{mazur,bdm,lempert,west,cajueiro}. However, there is more to the story.

In behavioral studies, the rationality of agents has long been a matter of discussion and debate \cite{kahneman}. In the studies of discounting, rational 
behavior of an agent is characeterized by inter-temporal consistency which may be understood from the following example. Let us 
assume that two sets of a pair of options are given to  participants. Options in {\it set 1} are  A1) USD 100 available after 365 days, and  B1) USD 120 available after 372 days. Options in {\it set 2} are A2) USD 100 available right now, and  B2) USD 120 available after 7 days. Choices A1 and A2 are more impulsive than B1 and B2 respectively because these choices manifest a preference for smaller-sooner rewards. However, things become interesting when a participant is asked to choose one option from
both sets during a single experiment. Many people may opt for the smaller-sooner reward in set 1, whereas for set 2 they settle in for a
larger-later reward (despite the waiting time in both cases are 7 days). This inconsistency in degree of impulsivity is known as `irrationality'. However, if in both cases one takes impulsive decisions, that is still rational, and the same is true for those who consistently opt for larger-later rewards in both cases.

In the case of rational agents, the exponential model of delay discounting is seen to be appropriate. The exponential delay discounting model considers an exponential dependence of the subjective value of a reward on delay given by,
\be
V(D)=V(0)\exp({-\kappa D}),
\label{expmodel}
\ee
where $V(D)$ is the discounted value of a reward after a delay $D$, $V(0)>V(D)$ is the undiscounted value of the reward, and $\kappa$
is the impulsivity parameter. 

Exponential model is related to the situation when the discount rate is constant over time. However, things change 
when this is not the case. In that case, the exponential model becomes unable to explain inter-temporal inconsistency described in a previous paragraph. Several other models like the hyperbolic model \cite{rachilinhyp} have been proposed to consider this phenomenon in the behavioral studies. But, the hyperbolic model cannot differentiate between inconsistency
and impulsivity. This is the reason why Takahashi \cite{takahashi1} proposed a delay discounting model based on the Tsallis $q$-exponential function which was also
used by Cajueiro \cite{cajueiro} in a study on inter-temporal choices. 

Unlike the exponential model, the Tsallis model is described by two parameters, $q$ (also called the Tsallis $q$ parameter) and impulsivity $\kappa$.
A fundamental approach involving the mathematical techniques of complex Hilbert space used in quantum mechanics has been taken by Yukalov and Sornette
\cite{yukalov}. By considering a time-dependent decay rate in the evolution of a `prospect set', they obtain a power-law time discounting function which is similar to the Tsallis function. Hence the Tsallis function comes both as a necessity to describe the observed phenomenon and as a consequence of a very general mathematical ritual. Because of these reasons, the Tsallis $q$-exponential function (henceforth to be called the Tsallis function) has been used in many behavioral studies over the years \cite{takahashi2,takahashiqdt,lempert,fpsy1,fph1,ejmbe}.



As indicated in one of them \cite{takahashi2}, the Tsallis function has its origin in physics. This function can be obtained from the Tsallis entropy, 
which is a generalization of the Boltzmann-Gibbs entropy.  While maximization of the 
Boltzmann-Gibbs entropy gives rise to the exponential function, maximization of the Tsallis entropy yields a Tsallis $q$-exponential function. 
This approach of getting the Tsallis function may be termed as the `thermodynamic approach' which is a standard procedure of obtaining the equilibrium
(maximum entropy state, to be understood as the state with the maximum disorderliness) thermodynamic quantities in systems considered in physics. 

However, in the present article, we do not consider the approach involving maximization in a social system (see for example \cite{cajueiro}). Here we rather try to examine
why the Tsallis function may appear in a social system from a perspective that is different from the ones presented in \cite{takahashi2,yukalov}. 
The approach taken by us is known as the so-called `super-statistics approach' \cite{beck,wilkprl} which helps one build another intuitive understanding of the model. This approach is based on the fact that social and physical systems contain fluctuations. Let us take the example of the temperature of a room. Even though we say that the `room temperature' is 30$^\text{o}$C (say), this may not reflect the reality that the temperature in one corner of the room may be slightly below (say 29.8$^\text{o}$C), and the other corner may be slightly above (say 31.2$^\text{o}$C)
than 30$^\text{o}$. So, where does this number 30 come from? This may be the average temperature of the room. If, for example, we have $n_1$ different points in the room having temperature $T_1$,  $n_2$ different points in the room having temperature $T_2$ and so on, the average 
temperature of the room is defined by $\left(n_1T_1+n_2T_2+....\right)/(n_1+n_2+...)$. In social systems too we encounter fluctuating quantities that follow some distributions, 
and the quantities like mean, variance become of our interest.

The novelty of the present work lies in the fact that from this simple consideration of a fluctuating system, the parameters used in the Tsallis model may be 
given some physical meaning. Although we do not deny that the Tsallis parameter (to be denoted as $q$) is already identified with the `measure of inconsistency' \cite{cajueiro,takahashi1},
our approach provides another intuitive way of looking at the model parameters. With this general consideration, we propose a model called the effective exponential
model which contains a Tsallis function and has some correspondence with the existing Tsallis models. 


The paper is organized in the following way. In the next section, we discuss the effective exponential model in detail. Section~\ref{discussion} contains a detailed discussion of our findings. 
Next, we summarize and conclude in Sec.~\ref{sumout}.

\section{Description of the model and its application}
\label{description}

\subsection{Fluctuating impulsivity: how is it distributed?}

\begin{figure*}[!htb]
\vspace*{+0cm}
\minipage{0.43\textwidth}
\includegraphics[width=\linewidth]{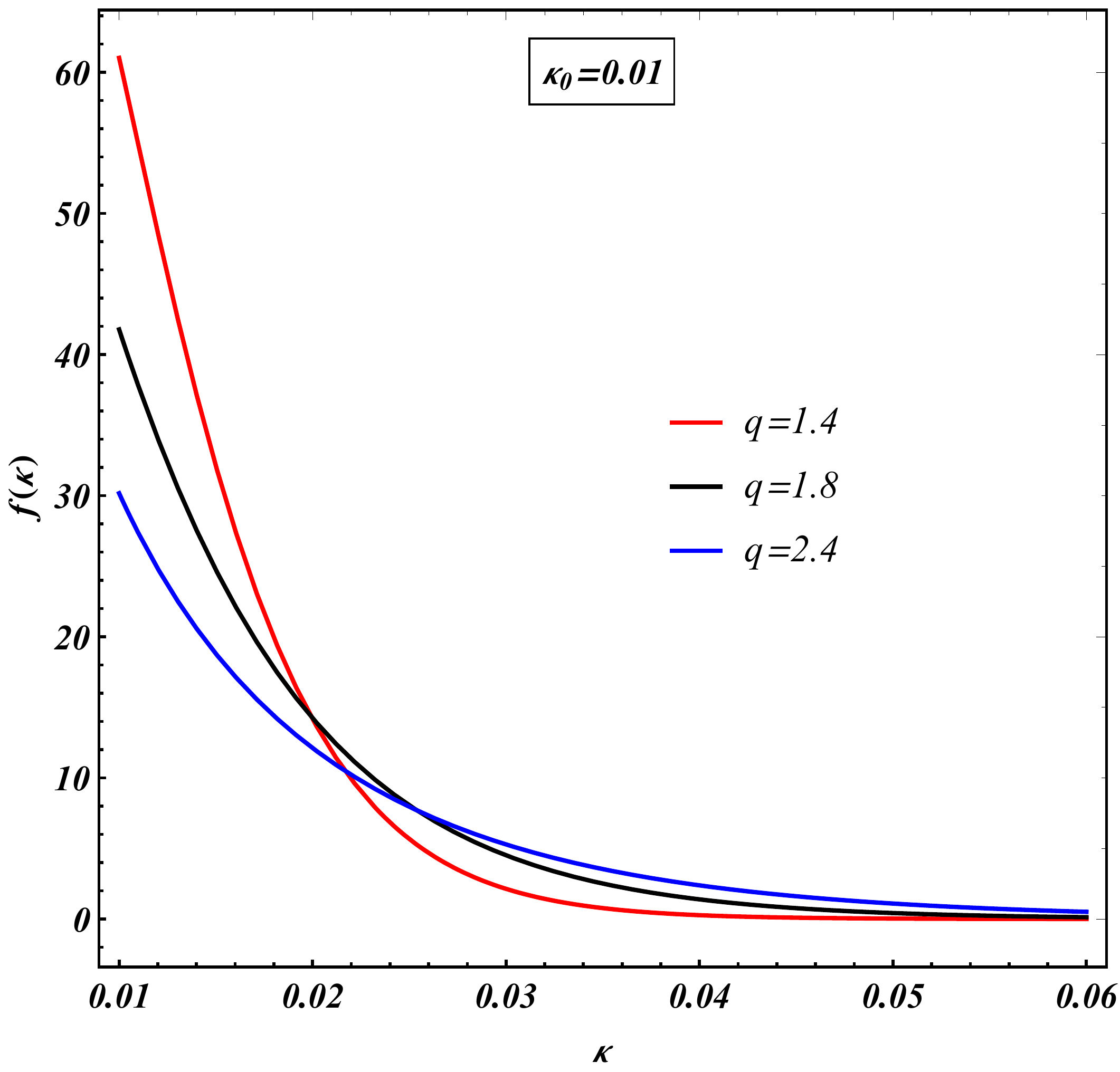}
\caption{Gamma distributions for a fixed $\kappa_0$ and different $q$ values.}
\label{figgamma1}
\endminipage\hfill
\minipage{0.43\textwidth}
\vspace*{+0cm}
\hspace*{-0.5cm}
\includegraphics[width=\linewidth]{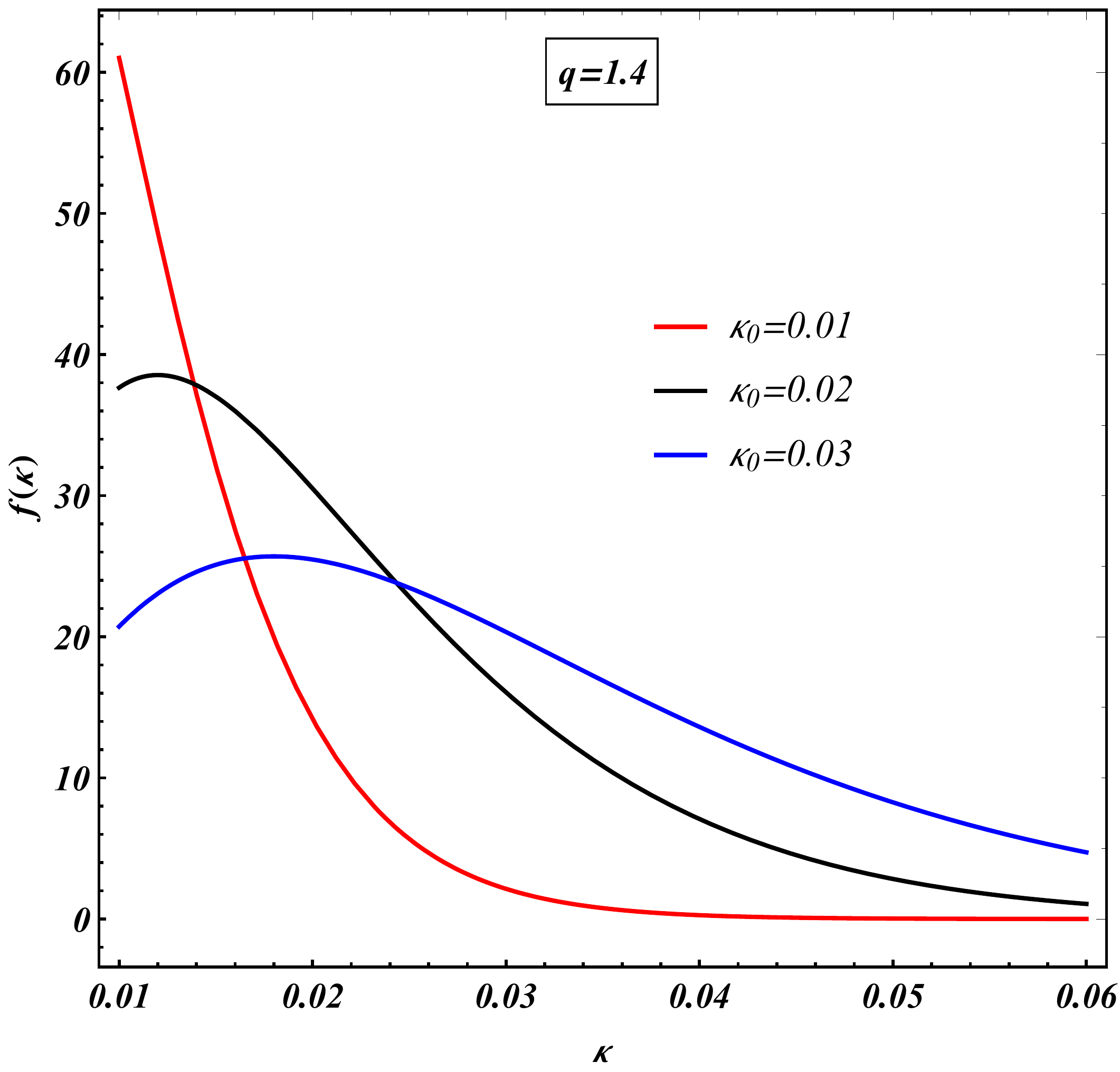}
\caption{Gamma distributions for a fixed $q$ and different $\kappa_0$ values.}
\label{figgamma2}
\endminipage\hfill
\end{figure*}

We consider a social system in which impulsivity $\kappa$ is a positive, random variable. But we assume that we do not know anything about how they are distributed. 
In a typical statistical problem, we have a random variable whose distribution is unknown. In that case, the gamma distribution serves as the least-informative option  \cite{hogg}. 
Let us assume that the impulsivity distribution of a sample is given by the normalized (integrates to 1) gamma distribution,
\be
f(\kappa) = \frac{1}{\Gamma\left(\frac{1}{q-1}\right)} \left\{\frac{1}{(q-1)\kappa_0}\right\} \left\{\frac{\kappa}{(q-1)\kappa_0}\right\}^{\frac{1}{q-1}-1} \exp\left(-\frac{\kappa}{(q-1)\kappa_0}\right),
\label{kappadist}
\ee
where $q>1$ and $\kappa_0>0$ (so that $f(\kappa)$ is a positive real number) are two parameters, and the $\Gamma$ function is represented by the following integral when $q>1$, 
\be
\Gamma \left(\frac{1}{q-1}\right) = \int_0^{\infty} \exp(-x) x^{\frac{1}{q-1}-1} dx.
\label{gammaintrep}
\ee
For different $q$ and $\kappa_0$ values the gamma distribution in Eq.~\eqref{kappadist} give rise to different shapes as can be seen in Figs.~\ref{figgamma1} and ~\ref{figgamma2}.  

\subsection{Impulsivity fluctuation in the exponential model and an effective factor}

We consider the exponential model of delay discounting given by Eq.~\eqref{expmodel} and consider that impulsivity $\kappa$ is a fluctuating quantity.
Considering that $\kappa$ in Eq.~\eqref{expmodel} is distributed according to Eq.~\eqref{kappadist}, we can define an `effective exponential factor' (essentially an average) 
defined by,

\bea
\epsilon_{\text{eff}} &=& \frac{\int_0^{\infty} d\kappa \exp({-\kappa D}) f(\kappa)}{\int_0^{\infty} d\kappa f(\kappa)} \nn\\
&=& \int_0^{\infty} d\kappa \exp({-\kappa D}) f(\kappa) 
\label{geff}
\eea
in which each $\kappa$ in the exponential is now weighted by the normalized distribution $f(\kappa)$.

The factor $\epsilon_{\text{eff}}$ accommodates fluctuating $\kappa$ values (which may range from very small to very large values) already present in a social system. In Eq.~\eqref{expmodel},
we replace the exponential function with $\epsilon_{\text{eff}}$, and propose the effective exponential model given by,
\be
V(D) = V(0) \epsilon_{\text{eff}}
\ee

Defining an effective exponential factor is the reason why the proposed model is named the `effective exponential model'. This approach is used in physical systems
\cite{wilkprl} where temperature is a fluctuating quantity and an effective Boltzmann factor is defined (Boltzmann is exponential). We show that this effective exponential factor 
is nothing but the Tsallis function. An analytical closed form of $\epsilon_{\text{eff}}$ is obtained as described below.

From Eq.~\eqref{kappadist} we have,
\bea
\epsilon_{\text{eff}} &=& \int_0^{\infty} d\kappa \exp({-\kappa D}) f(\kappa) \nn\\
&=& \frac{1}{\Gamma\left(\frac{1}{q-1}\right)} 
\left\{\frac{1}{(q-1)\kappa_0}\right\} \int_0^{\infty} d\kappa \exp({-\kappa D}) 
\left\{\frac{\kappa}{(q-1)\kappa_0}\right\}^{\frac{1}{q-1}-1} \exp\left(-\frac{\kappa}{(q-1)\kappa_0}\right) \nn\\
&=& \frac{1}{\Gamma\left(\frac{1}{q-1}\right)} 
\left\{\frac{1}{(q-1)\kappa_0}\right\}^{\frac{1}{q-1}} 
\int_0^{\infty} d\kappa \exp \left[{-\kappa \left(D+\frac{1}{(q-1)\kappa_0}\right)}\right]
\kappa^{\frac{1}{q-1}-1}
\eea
Now, defining $\kappa \left(D+\frac{1}{(q-1)\kappa_0}\right) =\kappa'$, the above integral can be written as,
\bea
\epsilon_{\text{eff}} &=& \frac{1}{\Gamma\left(\frac{1}{q-1}\right)} 
\left\{\frac{1}{(q-1)\kappa_0}\right\}^{\frac{1}{q-1}} 
\left(D+\frac{1}{(q-1)\kappa_0}\right)^{-\frac{1}{q-1}}
\int_0^{\infty} d \kappa' \exp(-\kappa') \kappa'^{\frac{1}{q-1}-1}.
\eea
From Eq.~\eqref{gammaintrep}, the integral in the above equation is nothing but $\Gamma\left(\frac{1}{q-1}\right)$ and hence,
\be
\epsilon_{\text{eff}} = \left(1+(q-1)\kappa_0 D\right)^{\frac{1}{1-q}} = \exp_q(-\kappa_0 D),
\label{qexpdef}
\ee
where $\exp_q$ is the $q$-exponential function \cite{Tsal88} which becomes the conventional exponential function in the limit $q\rightarrow1$. 
Hence, a modified form of Eq.~\eqref{expmodel} is given by,
\bea
V(D) &=& V(0) \exp_q(-\kappa_0 D) \nn\\
&=& \frac{V(0)} {\left(1+(q-1)\kappa_0 D\right)^{\frac{1}{q-1}}}. ~~~~~~\text{(--EEM--)} 
\label{ourTsmodel}
\eea
This is the main result of our paper which describes an `effective exponential model (EEM)' of delay discounting. Takahashi \cite{takahashi1}
has also proposed a similar Tsallis model. However, there are still some differences among the EEM and the other Tsallis model. 
We will discuss that in Section~\ref{discussion}.

\subsection{Significance of the parameters $q$ and $\kappa_0$}

Now, let us investigate what the parameters $\kappa_0$ and $q$ signify. As shown next, parameter $\kappa_0$ is the mean impulsivity of the system,
and $q$ is related to the variance of the impulsivity distribution. From Eq.~\eqref{kappadist},
\bea
\langle \kappa \rangle &=& \int_0^{\infty} d\kappa ~\kappa f(\kappa)  \nn\\
&=& \frac{1}{\Gamma\left(\frac{1}{q-1}\right)} 
\left\{\frac{1}{(q-1)\kappa_0}\right\}^{\frac{1}{q-1}}
\int_0^{\infty} d\kappa 
~\kappa^{\frac{1}{q-1}} \exp\left(-\frac{\kappa}{(q-1)\kappa_0}\right) 
\eea
By defining $k/(q-1)\kappa_0=\kappa''$ and using Eq.~\eqref{gammaintrep} we obtain,
\bea
\langle \kappa \rangle &=& \frac{\Gamma\left(\frac{1}{q-1}+1\right)}{\Gamma\left(\frac{1}{q-1}\right)} 
\times (q-1) \kappa_0 \nn\\
&=& \frac{\Gamma\left(\frac{1}{q-1}\right)}{(q-1)\Gamma\left(\frac{1}{q-1}\right)} 
\times (q-1) \kappa_0 \nn\\
&=& \kappa_0
\label{kappamean}
\eea

Now, to find out the variance, we need to find out the mean square impulsivity. Again, using Eq.~\eqref{kappadist} and redefining the integration variable appropriately
we obtain,
\bea
\langle \kappa^2 \rangle &=& \int_0^{\infty} d\kappa ~\kappa^2 f(\kappa)  \nn\\
&=& \frac{1}{\Gamma\left(\frac{1}{q-1}\right)} 
\left\{\frac{1}{(q-1)\kappa_0}\right\}^{\frac{1}{q-1}}
\int_0^{\infty} d\kappa 
~\kappa^{\frac{1}{q-1}+1} \exp\left(-\frac{\kappa}{(q-1)\kappa_0}\right) \nn\\
&=& \frac{\Gamma\left(\frac{1}{q-1}+2\right)}{\Gamma\left(\frac{1}{q-1}\right)} 
\times (q-1)^2 \kappa_0^2 \nn\\
&=& q \kappa_0^2
\eea
Hence, 
\bea
\frac{\langle \kappa^2 \rangle - \langle \kappa \rangle^2}{\langle \kappa \rangle^2} = q-1;~~
\Rightarrow q = 1+\frac{\langle \kappa^2 \rangle - \langle \kappa \rangle^2}{\langle \kappa \rangle^2}
\label{kappavar}
\eea
Eqs.~\eqref{kappamean} and \eqref{kappavar} show that the $\kappa_0$ parameter is the mean impulsivity of the
sample, and the $q$ parameter is related to the relative variance of impulsivity of the sample. 
From Eq.~\eqref{kappavar} it is evident that $q>1$. Hence, the $q\rightarrow1$ limit signifies a very narrow impulsivity 
distribution with negligible fluctuation. A similar relation for physical systems can be found in Ref.~\cite{wilkprl}
which involves temperature, and Ref.~\cite{wilkprc} which involves particle number.

%


\section{Discussion}
\label{discussion}
\subsection{Facets of impulsivity}
In this article, we consider fluctuations present in a system and we argue that the gamma distribution given by $f(\kappa)$ is the 
most general distribution of a fluctuating quantity. In that sense, our approach is based on a condition (fluctuation) 
that is ubiquitous in any system. In a social system, fluctuating impulsivity is a common feature. However, researchers 
always distinguish between two kinds of impulsivity, trait and state, also called `facets', denoted by $\kappa_\text{T}$ and $\kappa_{\text{S}}$ respectively. 
A trait characteristic is inherent, and a person (or a representative of any other species)  is born with it. On the other hand, the 
state characteristic develops over time and varies. Hence, when one talks about fluctuations, it may seem that she implies the 
state characteristics only. However, in a study \cite{traitstate} it has been found that that university soccer players with 
higher trait anxiety scores may experience increased state anxiety under pressure. Though it discusses a connection between 
trait and state anxiety, yet it is possible that the trait and state impulsivity are also connected. Hence we consider impulsivity obtained
from delay discounting data as a combination of both. That is because a more or less unaltered quantity combined with an alternating 
quantity gives rise to another alternating quantity. 

\subsection{Drift and diffusion of impulsivity}
It is interesting to note in Figs.~\ref{figgamma1} and
\ref{figgamma2} that when we vary $\kappa_0$ and $q$ of the distribution, the peak and the width of the distribution changes. This shift in the peak of average impulsivity and width is analogous to the velocity-space drift and diffusion of a collection of particles traversing inside a medium. In this case, the average velocity and the width change because of the interaction with the medium molecules. We can similarly think about the change in the impulsivity distribution in terms of {\it drift and diffusion in the impulsivity space}. If we study the time evolution of the impulsivity distribution $f(\kappa)$ of a test sample, it can schematically be written in terms of two terms representing the drift ($\gamma_{\kappa}>0$) and diffusion ($\delta_{\kappa}>0$) in the following way,
\bea
\dot{f}(\kappa) = -\text{change~due~to~}\gamma_{\kappa} + \text{change~due~to~}\delta_{\kappa},
\eea
where the dot on $f(\kappa)$ signifies total time derivative. Time evolution of the impulsivity distribution of a sample may be a consequence of many
factors like changing social status, social interaction and so on.

\subsection{Comparison of the EEM with another Tsallis-like model}
Another Tsallis-like model proposed by Takahashi in Ref.~\cite{takahashi1} has the following form according to our notations,
\bea
V(D) &=& \frac{V(0)}{\exp_q \left({\kappa_0 D}\right)} \nn\\ 
&=& \frac{V(0)} {(1+(1-q)\kappa_0 D)^{\frac{1}{1-q}}}.  ~~~~~~\text{(--TM--)}
\label{takTsmodel}
\eea
Eq.~\eqref{takTsmodel} is to be referred to as the Takahashi model (TM) from now on. If $q$ is replaced by $2-q$ in the TM, one obtains the EEM (Eq.~\ref{ourTsmodel}), and this
is the reason why the TM and EEM may be termed as `dual to each other'. This duality reminds one of the same correspondence existing between the $q$-exponential function and its
inverse \cite{eqinv}. The EEM relates $q$ to the relative variance of the impulsivity distribution, and it is always positive. However, the $q$ values in the TM model can assume negative values. In spite of the difference in their appearances, it is seen that numerically these two models are equivalent. However, the parameter values in two approaches have a $q\leftrightarrow2-q$ correspondence because of the dual nature.

There is another subtle difference between the two Tsallis-inspired models which we discuss now. This discussion is based on the fact that we demand $V$ to be a real-valued quantity and not a complex number. From Eq.~\eqref{ourTsmodel} it is clear that if $1+(q-1)\kappa_0 D$ is negative, one may get a complex valued quantity because the $q$ parameter (also called the Tsallis parameter or the entropic index) is a real number. For example, for $q=0.1$, $\kappa_0=0.2$ and $D=10$, $1+(q-1)\kappa_0 D = -0.8$ and $\left(1+(q-1)\kappa_0 D\right)^{1/(1-q)}$ is a complex number. However, since we have $q>1,~\kappa_0>0$ and $D>0$ in our model (Eq.~\ref{ourTsmodel}), we do not encounter
complex values. The same constraint of positivity of the argument leads to an upper-bound of the $q$ values in the TM,
\be
q<1+\frac{1}{\kappa_0 D}.
\ee

\subsection{Different limits of the EEM}

Now, let us discuss different limits of the effective exponential model. In the limit $q\rightarrow 1$ (vanishing fluctuation), Eq.~\eqref{ourTsmodel} gives back the exponential
model of delay discounting as shown below,
\bea
\lim_{q\rightarrow1} V(D) &=& \lim_{q\rightarrow1}  V(0) \left(1+(q-1)\kappa_0 D\right)^{-\frac{1}{q-1}} \nn\\
&=& \lim_{q\rightarrow1}  V(0) \exp \left({-\frac{1}{q-1}\ln \left(1+(q-1)\kappa_0 D\right)}\right) \nn\\
&=& \lim_{q\rightarrow1}  V(0) \exp \left({-\frac{1}{q-1} \left\{(q-1)\kappa_0 D - \mathcal{O}(q-1)^2+...\right\}}\right) \nn\\
&=& V(0) \exp\left(-\kappa_0 D\right)
\label{ourTsmodelq1lim}
\eea
When we take the $\kappa_0D\rightarrow0$ limit, Eq.~\eqref{ourTsmodelq1lim} gives back the hyperbolic equation given by,
\be
\lim_{\kappa_0D\rightarrow0} V(D) \approx \frac{V(0)}{1+\kappa_0D}.
\ee
Also, when $q=2$, Eq.~\eqref{ourTsmodel} gives back the hyperbolic model. Hence the EEM generalizes the existing models of delay discounting which may be 
obtained by taking appropriate limits of the EEM.


\section{Summary, conclusions, and outlook}
\label{sumout}
To summarize, considering impulsivity to be a fluctuating quantity we describe an effective exponential model of delay
discounting phenomenon. This model described in the article is found to be dual to the Tsallis model of delay discounting 
proposed by Takahashi \cite{takahashi1}. In that sense, our findings support the existing 
Tsallis models which efficiently describe different aspects of delay discounting (like the inter-temporal inconsistency) which
are important in deciding policies in many sectors like finance, medical etc. 

The novelty of our article lies in the approach it takes. We consider a very general characteristic of a system (fluctuation) and 
propose a modification of the existing exponential model to arrive at another intuitive understanding of the Tsallis model. 
With the help of this approach we explain the significance of the model parameters. We find that 
\\
\begin{itemize}
\item the Tsallis parameter $q$ is related to the relative variance (variance divided by average square) of the impulsivity distribution
and is greater than 1.
\item the impulsivity parameter $\kappa_0$ is actually the mean of the impulsivity distribution.
\end{itemize}
It will be interesting to apply this model for participants from varied age groups and social backgrounds to study
how impulsivity changes. It will also be interesting to see how the personality of an individual affects this behavior. From the observations, 
one may parameterize the evolution the $q$ and $\kappa_0$ parameters with age to predict the impulsivity behavior of people from certain 
age group from a certain socio-cultural background. These findings will be important in deciding policies in relevant areas. 

Our study can throw light upon some other interesting aspects of social science in general. We try to elaborate on this in the next few sentences.
As mentioned in Section~\ref{discussion}, we consider impulsivity obtained from delay discounting data to be a combination of trait and state facets.
However, how to combine these two facets to get the combination is another interesting question which future researchers may revisit. In many studies it is
assumed to be a sum score given by the addition of facets. Since we consider impulsivity to be bi-faceted, the total impulsivity may be given
by the sum score of $\kappa_\text{T}$ and $\kappa_{\text{S}}$. In that case, one immediately finds that addition is only one of the options among
a large number of operators which combine these two numbers. There can also be some other ways to combine the facets. One may propose a {\it gedankenexperiment}
\cite{gedanken} in which the trait and state impulsivity of a group of participants are quantified from questionnaires, and the combined impulsivity 
of the same participants are measured by fitting the data obtained from the delay-discounting experiments using the Tsallis model. 
From these three sets of numbers, one can try to find out a rule of combining them. This question becomes really important when one needs
to score competitors for providing grants, or for providing some aid and in many other situations in which a questionable method of evaluation
may be disappointing (for the marginalized section, for example). 

\textcolor{black} 
{We would like to point out that the superstatistics approach may also be taken in describing financial market data using the
$q$-gaussian distribution \cite{tsfinnewjphys,tsfinphysrept} that is obtained by replacing the exponential function in Eq.~\eqref{geff} by a gaussian. In Refs.~\cite{tsfinnewjphys,tsfinphysrept}, authors observe that $q$ approaches unity when the time lag increases for the financial return distribution. This can be explained if $q$ can be identified as an internal variable that follows a relaxation-like equation \cite{qint},
\begin{equation}
\frac{dq}{d\delta t} = -\frac{q-q_{\text{eq}}}{\tau},
\end{equation}
where $q_{\text{eq}}=1$, $\delta t$ is the time-lag, and $\tau$ is a characteristic time scale. The solution of the above equation yields $q(\delta t)=1+(q_{\text{eq}}-1)e^{-\frac{\delta t}{\tau}}$, and supports the observation.}

To conclude, the effective exponential model lends support to the Tsallis model of delay discounting. However, its implications may go way beyond
that as discussed in the article. We have provided some insight into the model, and have raised some questions that seemed to be relevant
for a deeper understanding. With this input, we hope that future studies will shine more light on this very interesting phenomenon of delay discounting.



\begin{thebibliography}{100}
\bibitem{eqinv} $e_q(x)e_{2-q}(-x)=1$ as can be seen from Eq.~\eqref{qexpdef}.
\end{thebibliography}
\end{document}